\documentclass[structabstract]{aa}
\usepackage{graphicx}
\usepackage{natbib}
\usepackage{txfonts}
\usepackage{upgreek}
\usepackage{url}

\begin{document}

\title{Mid-infrared interferometry\\
  of the massive young stellar object NGC\,3603 - IRS\,9A
  \thanks{Based in part on observations collected at the European
  Southern Observatory, Chile (Prop. No. 074.C-0062) and the
  Gemini South Observatory, Chile.}}

\author{
  S.~Vehoff\inst{1,2}         \and 
  C.A.~Hummel\inst{3}         \and 
  J.D.~Monnier\inst{4}        \and  
  P.~Tuthill\inst{5}          \and
  D.E.A.~N\"urnberger\inst{2} \and
  R.~Siebenmorgen\inst{3}     \and
  O.~Chesneau\inst{6}         \and
  W.J.~Duschl\inst{7,8}
}

\institute{
  Institut f\"ur Theoretische Astrophysik, Zentrum f\"ur Astronomie der
  Universit\"at Heidelberg, Albert-Ueberle-Str.~2, 69120 Heidelberg,
  Germany
  \and
  European Organisation for Astronomical Research in the Southern Hemisphere,
  Casilla 19001, Santiago 19, Chile 
  \and
  European Organization for Astronomical Research in the Southern Hemisphere,
  Karl-Schwarzschild-Str.~2, 85748 Garching bei M\"unchen, Germany 
  \thanks{Correspondence: {\tt chummel@eso.org}}
  \and
  Department of Astronomy, University of Michigan, Ann Arbor, MI 48109, USA
  \and
  Sydney Institute for Astronomy, School of Physics, University of Sydney,
  NSW 2006, Australia
  \and
  Observatoire de la C\^ote d'Azur, Dpt. Gemini-CNRS-UMR 6203, Avenue Copernic,
  06130 Grasse, France
  \and
  Institut f\"ur Theoretische Physik und Astrophysik der
  Christian-Albrechts-Universit\"at zu Kiel, Leibnizstr. 15, 24118 Kiel, Germany
  \and
  Steward Observatory, The University of Arizona, 933 N. Cherry Ave., Tucson,
  AZ 85721, USA
}

\date{Received: 26 October 2009; accepted: 7 July 2010}  

\abstract
{Very few massive young stellar objects (MYSO) have been studied in
the infrared at high angular resolution due to their rarity and large
associated extinction. We present observations and models for one of
these MYSO candidates, NGC\,3603 IRS\,9A.}
{Our goal is to investigate with infrared interferometry the structure
of IRS\,9A on scales as small as 200\,AU, exploiting the fact that a
cluster of O and B stars has blown away much of the obscuring foreground
dust and gas.}
{Observations in the N-band were carried out with the MIDI beam combiner
attached to the VLTI, providing spatial information on scales of about
25\,--\,95 milli-arcseconds (mas). Additional interferometric observations
which probe the structure of IRS\,9A on larger scales were performed
with an aperture mask installed in the T-ReCS instrument of Gemini South.
The spectral energy distribution (SED) is constrained by the MIDI N-band
spectrum and by data from the Spitzer Space Telescope.  Our efforts to
model the structure and SED of IRS\,9A range from simple geometrical models
of the brightness distribution to one- and two-dimensional radiative
transfer computations.}
{The target is resolved by T-ReCS, with an equivalent (elliptical)
Gaussian width of 330\,mas by 280\,mas (2300 AU by 2000 AU). Despite
this fact, a warm compact unresolved component was detected by MIDI
which is possibly associated with the inner regions of a flattened
dust distribution.  Based on our interferometric data, no sign of
multiplicity was found on scales between about 200\,AU and 700\,AU
projected separation. A geometric model consisting of a warm (1000 K) ring
(400 AU diameter) and a cool (140 K) large envelope provides a good fit
to the data.  No single model fitting all visibility and photometric data
could be found, with disk models performing better than spherical models.}
{While the data are clearly inconsistent with a spherical dust distribution 
they are insufficient to prove the existence of a disk but rather hint
at a more complex dust distribution.}

\keywords{techniques: interferometric - stars: circumstellar matter - stars:
          early-type - stars: formation - stars: pre-main sequence - stars:
          individual: NGC\,3603 IRS\,9A}

\titlerunning{Mid-infrared interferometry of NGC\,3603 IRS\,9A}

\authorrunning{S.~Vehoff et al.}

\maketitle

\section{Introduction} \label{introduction}

The formation and evolution of massive stars is so rapid that by
the time the star has reached the zero-age main sequence, it is still
enshrouded in large amounts of gas and dust of the molecular cloud from
which it was born. Thus, high extinction is usually present at this
stage and prevents the optical identification and classification of
the star, as well as a more detailed study of the inner regions of the
circumstellar dust. Furthermore, high-mass stars are rare objects and
hence generally much more distant than low-mass star forming regions,
requiring high angular resolution for study. For a detailed account of
the theoretical and observational issues connected to the formation of
(high-mass) stars we refer to the review articles by \citet{mckee2007}
and \citet{zinnecker2007}.

In the case of formation and early evolution of low-mass (T\,Tauri)
stars and intermediate-mass (Herbig Ae/Be) stars, the longer
timescales involved allow us to study the mass accretion process,
the chemical evolution of the dust in the regions close to the star,
and the subsequent dispersion of the circumstellar material in great
detail. A key ingredient of current models is the formation of a disk
which plays an important role in removing angular momentum from the
infalling material. Thus, one would expect a similar scenario for the
formation of high-mass stars, even more so since the problem of their
extreme radiation pressure acting upon the spherical dust envelopes would
limit their final mass \citep[e.g.,][]{wolfire1987,krumholz2009}. But it
appears that one cannot simply scale the low-mass star formation models
to higher masses \citep{zinnecker2007}, since, amongst other reasons,
most, if not all, high-mass stars are born in cluster environments
\citep[e.g.,][]{dewit2005} where accretion might occur in a competitive
manner \citep{bonnell1997,bonnell2001,2008ASPC..387..208C}.

The target of our study, \object{[FPA77] NGC 3603 IRS 9A},
has been identified as one of
the most luminous mid-infrared sources located within the giant
H\,{\footnotesize II} region NGC\,3603 \citep{2003A&A...404..255N}. This
star forming region is dominated by one of the densest and most
massive clusters of O and B type main sequence stars in the Galaxy
\citep{moffat1994}.  The stellar winds and ionising radiation
of these stars have blown away most of the gas and dust of the
molecular cloud core around IRS\,9A \citep{2008JPhCS.131a2025N},
which lies at a projected distance of only 2.5\,pc from the central
cluster.  In spite of a distance towards NGC\,3603 of about ($7 \pm
1$)\,kpc, the visual foreground extinction towards the cluster is only
4\,--\,5\,mag \citep{sher1965,moffat1983,melnick1989}. Using infrared
colour-magnitude diagrams and pre-main sequence evolutionary tracks,
\citet{2003A&A...404..255N} was able to deduce a mass of $\approx
40\,M_{\sun}$ for IRS\,9A. Its luminosity was estimated to about $2.3
\times 10^5\,L_{\sun}$ and its age to $\approx 10^4$\,yr. The spectral
index $\alpha_{2.2-10\upmu{\rm m}} = 1.37$ is similar to those of
(low-mass) IR class I objects, indicating that IRS\,9A is still
embedded in significant amounts of gravitationally bound material.

These properties all argue for IRS\,9A's classification as a high-mass
young stellar object which, due to its advantageous location, can be
studied more easily. Observations at very high-angular resolution in
the mid-infrared hold the potential to yield valuable constraints
for the theoretical understanding of high-mass star formation,
including multiplicity, but only very few other MYSOs have been
studied via (mid-) infrared interferometry so far. Among them are W33A
\citep{dewit2007}, M8E-IR \citep{2009A&A...505..655L}, and
MWC\,297 \citep{acke2008}. We will compare the properties of IRS\,9A to
these objects later on in this work.

The remaining sections of this paper are organised as follows. In
Sect.~\ref{observations} we present the main results of our observations,
shortly discussing the data reduction processes and immediate implications
of the individual results. In Sect.~\ref{analysisandmodels} we model the
circumstellar structure of IRS\,9A using simple geometrical models and
dedicated radiative transfer models.  Finally, we discuss the results
of these efforts in Sect.~\ref{discussion} and present our conclusions
in Sect.~\ref{conclusions}.

\section{Observations and data reduction} \label{observations}

\subsection{MIDI} \label{obs_midi}
Even though the MID-infrared Interferometric instrument (MIDI) on the VLTI is
primarily an interferometric instrument, photometry and imaging are by-products
of each observation. Thus, a total flux spectrum can be obtained as well as
images which are taken during the target acquisition procedure. Due to the use
of an adaptive optics system (MACAO) at the Coud\'{e} foci of each UT of the
VLT, these images can be analysed for medium scale structures of the target.
In the following, we will describe each of these three data products.

\subsubsection{Interferometry} \label{obs_midi_int}

\begin{figure}
  \includegraphics[width=1.0\columnwidth]{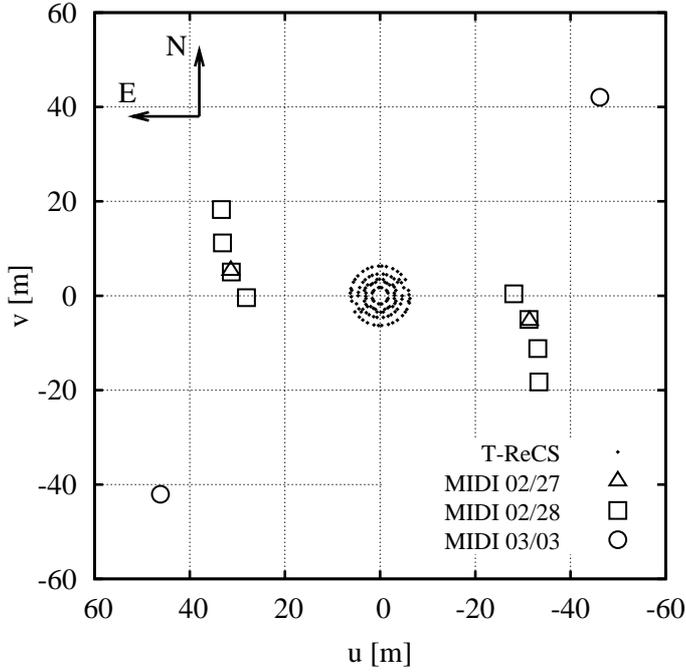}
  \caption{$(u,v)$-plane coverage of the interferometric observations with
    MIDI at the VLTI and the aperture masking observations with T-ReCS at
    Gemini South.}
  \label{f_MIDI_Gemini_uv}
\end{figure}

The interferometric observations using MIDI
\citep{leinert2003spie,leinert2003messenger} were carried out in February
and March of 2005.  The most important details of these observations
are summarised in Table~\ref{t_midilog}.  The resulting $(u,v)$-coverage is
shown in Fig.~\ref{f_MIDI_Gemini_uv} together with the coverage resulting
from the  T-ReCS observations. The synthesized interferometric beam
size is about 28 mas by 16 mas, PA of 47 degrees.  The observations of
NGC\,3603 IRS\,9A were interleaved with measurements of HD\,107446, our
calibrator, for which we adopted a diameter of 4.7\,mas based on the (V-K)
color index \citep{mozurkewich2003}. A star with this diameter is nearly
unresolved on a 40\,m baseline and at a wavelength of 8\,$\upmu$m ($V^2
= 0.97$). HD\,107446's angular distance to IRS\,9A is about 8$\degr$;
it has a spectral type of K3.5III and an IRAS flux of 32.4\,Jy in the
12\,$\upmu$m filter. MIDI was used with the high-sensitivity beam combiner
configuration (just two interferometric outputs) and the prism for
spectral dispersion, providing a spectral resolution of $R \approx 35$.

\begin{table}
\caption[]{Log of observations of IRS\,9A with MIDI, all performed in 2005. 
    The spatial resolution $\lambda/B_{\rm p}$ is calculated for 8\,$\upmu$m.
    BLPA = baseline position angle projected on sky (east of north), AM =
    airmass. The letters (a)--(d) refer to Fig.~\ref{f_MIA_V1}.}
\label{t_midilog}
\centering
\begin{tabular}{cccccr@{.}lc}
\hline\hline
Date   & Time    & Target     & $B_{\rm p}$ & $\lambda/B_{\rm p}$ & \multicolumn{2}{c}{BLPA}      & AM \\
       & (UT)    &            & [m]         & [mas]               & \multicolumn{2}{c}{[$\degr$]} &    \\ \hline
Feb 27 & 08h 01m & HD\,107446 & 38.0        & 43.4                &  60 & 6                       & 1.3\\
       & 08h 35m &    IRS\,9A & 31.8        & 52.0                &  79 & 6                       & 1.5\\
       & 09h 16m & HD\,107446 & 33.6        & 49.1                &  74 & 0                       & 1.4\\

Feb 28 & 06h 24m & HD\,107446 & 41.8        & 39.5                &  44 & 3                       & 1.2\\
(a)    & 06h 48m &    IRS\,9A & 38.0        & 43.5                &  60 & 6                       & 1.3\\
       & 07h 09m & HD\,107446 & 40.2        & 41.0                &  52 & 2                       & 1.2\\
(b)    & 07h 43m &    IRS\,9A & 34.9        & 47.3                &  70 & 6                       & 1.4\\
       & 08h 04m & HD\,107446 & 37.6        & 43.8                &  61 & 8                       & 1.3\\
(c)    & 08h 33m &    IRS\,9A & 31.6        & 52.2                &  80 & 0                       & 1.5\\
       & 08h 54m & HD\,107446 & 34.8        & 47.5                &  70 & 7                       & 1.4\\
(d)    & 09h 21m &    IRS\,9A & 28.0        & 58.9                &  89 & 9                       & 1.7\\
       & 09h 48m & HD\,107446 & 31.0        & 53.2                &  80 & 8                       & 1.5\\
Mar 3  & 04h 21m & HD\,107446 & 56.2        & 29.5                &  81 & 4                       & 1.3\\
       & 05h 27m & HD\,107446 & 59.4        & 27.8                &  95 & 8                       & 1.3\\
       & 07h 10m &    IRS\,9A & 62.5        & 26.5                & 131 & 9                       & 1.3\\
       & 07h 49m & HD\,107446 & 62.5        & 26.5                & 126 & 0                       & 1.3\\
       & 08h 54m &    IRS\,9A & 61.9        & 26.7                & 155 & 5                       & 1.6\\
       & 09h 16m & HD\,107446 & 62.2        & 26.5                & 145 & 2                       & 1.5\\
       & 09h 33m & HD\,107446 & 62.0        & 26.6                & 149 & 1                       & 1.5\\
       & 09h 51m & HD\,107446 & 61.9        & 26.7                & 153 & 3                       & 1.6\\
    \hline
  \end{tabular}
\end{table}

\begin{figure}
\includegraphics[width=1.0\columnwidth]{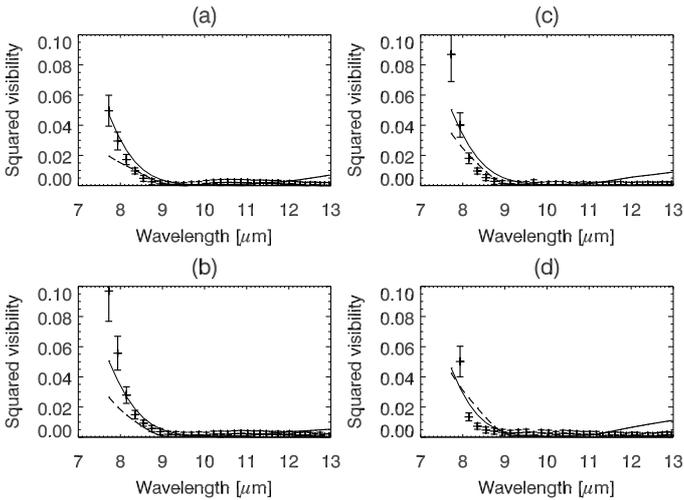}
\caption{Squared visibilities measured with MIDI on the UT2-UT3 baseline on
    Feb. 28, 2005, with letters (a)--(d) corresponding to the four observations
    listed in Table~\ref{t_midilog}. The solid lines are best fits of the
    geometrical model described in Section~\ref{geometricalmodels}, and
    the dashed lines are from the physical model described in
    Section~\ref{Diskmodels}.}
\label{f_MIA_V1}
\end{figure}

\begin{figure}
\includegraphics[width=1.0\columnwidth]{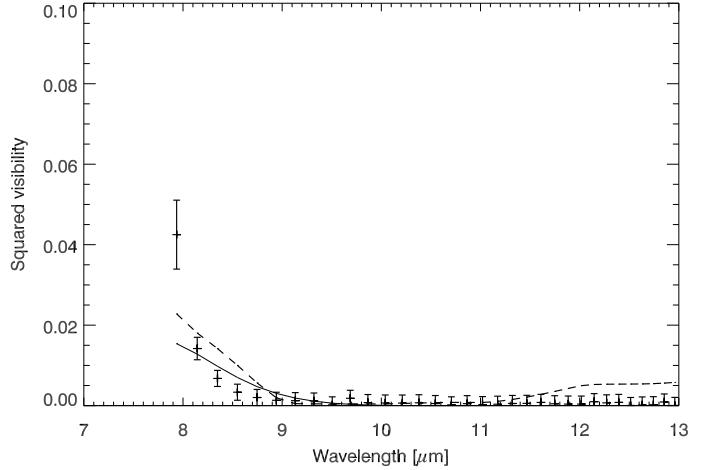}
\caption{Squared visibilities measured with MIDI on the UT3-UT4 baseline (first
    observation on Mar 3, 2005). The solid line is a best fit obtained from
    the geometrical model described in Section~\ref{geometricalmodels},
    and the dashed line is from the physical model described in
    Section~\ref{Diskmodels}.}
\label{f_MIA_V2}
\end{figure}

The data were reduced using the MIA+EWS (version 1.5) software
package\footnote{Available at:
\url{http://www.strw.leidenuniv.nl/~nevec/MIDI/index.html}}
which is described in detail in \citet{ratzka2005}. In order to
calculate the visibility, the correlated flux is normalised with the
total flux. The latter is obtained directly after the interferometric
measurements, using the individual telescopes and chopping between
the target position and a sky position (for IRS\,9A: position angle
$= 117\degr$, throw $= 17 \arcsec$).  The calibration of the target
visibilities was achieved by division with the calibrator visibilities,
interpolated for each epoch of the target observations. We estimated the
calibration uncertainty from the scatter of several measurements of the
calibrator on Feb.\ 28 to range from about 10\% at 13\,$\upmu$m to about
20\% at 8\,$\upmu$m.  The resulting calibrated visibilities of IRS\,9A are
shown in Fig.~\ref{f_MIA_V1} (UT2-UT3 baseline) and in Fig.~\ref{f_MIA_V2}
(UT3-UT4 baseline). The single visibility spectrum of Feb.\ 27 is not
shown as it is redundant, and the second observation of Mar.\ 3 was
taken at high airmass and did not yield data of good quality.
The data are made available at 
OLBIN\footnote{\url{http://olbin.jpl.nasa.gov/data/}}.

The visibilities are very similar for all our baselines. They drop to virtually
zero for wavelengths larger than about 9\,$\upmu$m where IRS\,9A appears fully
resolved by MIDI. Yet we detect a compact component (roughly 60 mas in diameter)
which emerges below this wavelength, causing a steep rise of the visibilities. 

\subsubsection{Spectro-photometry} \label{obs_midi_spec}

\begin{figure}
\includegraphics[width=1.0\columnwidth]{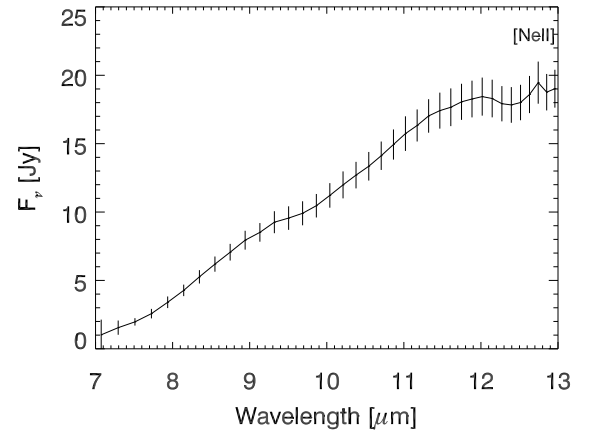}
\caption{Average spectrum of IRS\,9A, as obtained from the MIDI data taken
    on Feb 28, 2005. The small spike at 12.8\,$\upmu$m coincides with the
    [Ne\,{\footnotesize II}] line.}
\label{f_EWS_spectrum}
\end{figure}

Since HD\,107446 is not a spectro-photometric calibrator, we adopted for its
spectrum the template of HD\,163588 (K2III) from the list of spectro-photometric
standards by \citet{cohen1999}, and used the ratio of the IRAS 12\,$\upmu$m
fluxes to determine the relative scale.
The resulting flux-calibrated spectrum of IRS\,9A as measured with MIDI is
shown in Fig.~\ref{f_EWS_spectrum}. We averaged only spectra from Feb.\ 28,
2005 as they represent the largest homogeneous data set. The error bars are
based on the standard deviation of the spectra.

There are several interesting aspects concerning this spectrum. First, the
flux of IRS\,9A rises steeply with increasing wavelength.
Second, there is no sign of a silicate feature at about 9.7\,$\upmu$m,
neither in absorption nor in emission. This is rather uncommon for
dust-enshrouded objects like YSOs. Third, despite the low spectral
resolution of MIDI, there is a hint of the [Ne\,{\footnotesize II}]
emission line at 12.8\,$\upmu$m, prominently seen by Spitzer (see
Sect.~\ref{obs_Spitzer}).

\subsubsection{Imaging}

The chopped acquisition images of IRS\,9A were taken with the N8.7
filter ($\lambda_{\rm c} = 8.64\,\upmu$m, $\lambda_{\rm FWHM} =
1.54\,\upmu$m) and have a spatial resolution of about 0.3$\arcsec$.
Using the calibrator HD\,107446 as a PSF reference, the images can be
deconvolved to further enhance the resolution, using the procedure
described by \citet{2005A&A...435..563C}. We used the Lucy-Richardson
algorithm \citep{lucy1974} for this purpose, taken from the IDL Astronomy
Library\footnote{\url{http://idlastro.gsfc.nasa.gov/}}. However,
we detected a residual jitter in our images caused by a delay in the
acquisition by the adaptive optics loop each time the chop position
changed.  Simply re-centering the frames did not lead to useful results
due to distorted profiles while MACAO was closing the loop.  Therefore,
we only stacked the chopped frames which were well centered (about half
of the frames).

\begin{figure}
  \includegraphics[width=1.0\columnwidth]{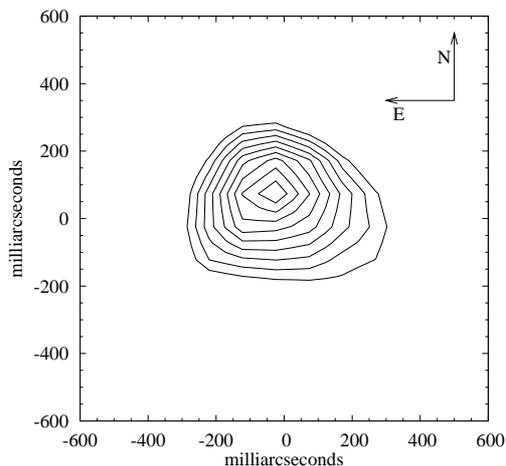}
  \caption{Deconvolved MIDI acquisition image of IRS\,9A in the N8.7 filter
    after 25 iterations. The contour lines are equally spaced for the fourth
    root of the intensity and lie at 3.7, 6.3, 10, 15, 22, 32, 44, 59 and 77 
    percent of the maximum intensity.}
  \label{f_acq}
\end{figure}

We stopped the algorithm after 25 iterations, a compromise between
image convergence and fitting noise. A representative example of our
deconvolved acquisition images is shown in Fig.~\ref{f_acq}. We performed
two-dimensional Gaussian fits to the deconvolved images in order to
determine the FWHM sizes and the orientation of the major axis. The
results are summarised in Table~\ref{t_acq} which lists the mean single
night values, and also the mean value of all our observations, using
the number of observations per night as the associated weight (N$_{\rm
obs}$). We did not use the images of the beam B of MIDI (UT3 on Feb.\
27 and Feb.\ 28, UT4 on Mar.\ 3) due to their slightly inferior image
quality and the associated systematically  larger FWHM values. Overall,
IRS\,9A has a Gaussian FWHM of about 389\,mas by 354\,mas, and a position
angle of the major axis of about 75 degrees (east of north). Despite the
different values of the parallactic angle of the observations, i.e. the
direction to north on the detector, ranging from 130 to 173 degrees, the
derived PA of the fitted ellipse was unchanged. Representing the overall
shape of the source, the fitting of the PA was apparently dominated by
the SW ridge in the source structure.

\begin{table}
  \caption{Results of the Gaussian fits to the deconvolved MIDI acquisition
    images.}
  \centering
  \setlength{\tabcolsep}{1.2ex}
  \begin{tabular}{ccccccc}
    \hline\hline
      Date           & Beam & $N_{\rm obs}$ & Major & Minor & PA  \\
                     &      &               & [mas] & [mas] & [$\degr$] \\ \hline
      Feb.\ 27, 2005 & UT2  &  1            & 402   &  339  & 82  \\
      Feb.\ 28, 2005 & UT2  &  4            & 392   &  366  & 71  \\
      Mar.\ 03, 2005 & UT3  &  2            & 375   &  339  & 81  \\
      Mean values    &      &  7            & 389   &  354  & 75  \\\hline
  \end{tabular}
  \label{t_acq}
\end{table}

\subsection{T-ReCS} \label{obs_Gemini}

We also observed NGC\,3603 IRS\,9A with the T-ReCS (Thermal-Region
Camera Spectrograph, \citet{debuizer2005}) instrument at the Gemini South
Observatory on January 2, 2005. A 7-hole non-redundant aperture mask was
used to reduce the weight of the short spacings realised within the 8\,m
aperture of the telescope, and therefore to yield high quality images at
the diffraction limit \citep[for a description of the aperture masking
technique see, e.g.,][]{monnier1999,tuthill2000}. The data were taken
with the Si-5 filter ($\lambda_{\rm c} \pm {\scriptstyle \Delta} \lambda
= 11.66\,\upmu{\rm m} \pm 0.57\,\upmu$m). The resulting $(u,v)$-coverage
is shown in Fig.~\ref{f_MIDI_Gemini_uv}; the synthesized beam is almost
circular with a diameter of 264 mas.

\begin{figure}
  \includegraphics[width=1.0\columnwidth]{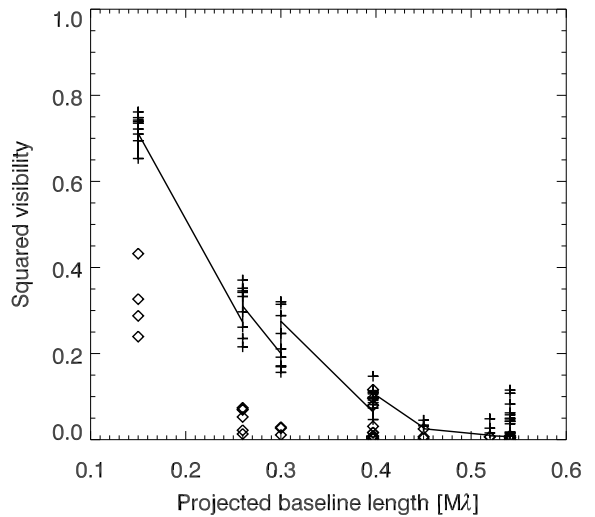}
  \caption{Squared visibility amplitudes from T-ReCS plotted versus projected
    baseline length. Error bars have been omitted for clarity; they are shown
    in Fig.~\ref{f_Gemini_blpa_1}. The vertical scatter of the visibility 
    amplitudes
    arises from non-axisymmetric structure, which causes some azimuthal
    dependence of the amplitudes. The solid line is a best fit obtained from
    the geometrical model described in Section~\ref{geometricalmodels},
    and the diamonds correspond to the physical model described in
    Section~\ref{Diskmodels}.}
  \label{f_Gemini_uvr}
\end{figure}

\begin{figure}
  \includegraphics[width=1.0\columnwidth]{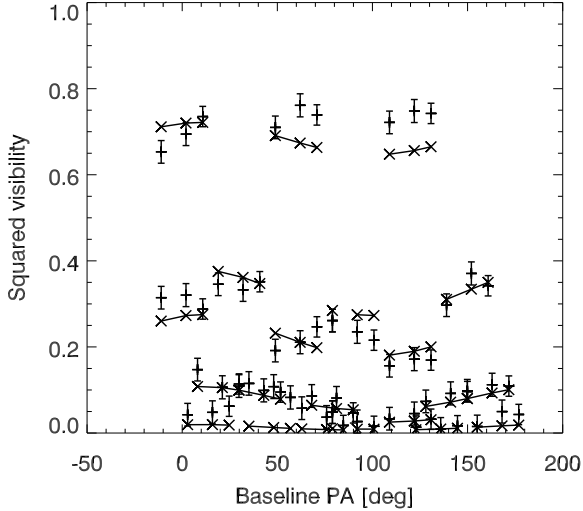}
  \caption{Squared visibility amplitudes from T-ReCS ($+$-symbols)
    plotted versus baseline
    position angle (0-180) The vertical offsets of the visibility amplitudes
    is due to the different lengths of baselines with the same position angle.
    The solid lines (with x-symbols)
    are the best fits obtained from the geometrical model described in
    Section~\ref{geometricalmodels}.}
  \label{f_Gemini_blpa_1}
\end{figure}

The interferograms, which result from the superposition of
all interference patterns between any combination of apertures in the mask,
are Fourier transformed to measure the visibility amplitudes and phases.
This is done for both target and calibrator, and the calibration procedure is
similar to the one used in long baseline interferometry. The resulting
squared visibility amplitudes are shown in Figs.~\ref{f_Gemini_uvr} and
\ref{f_Gemini_blpa_1}.
The data are made available at
OLBIN\footnote{\url{http://olbin.jpl.nasa.gov/data/}}.

\begin{figure}
\includegraphics[width=1.0\columnwidth]{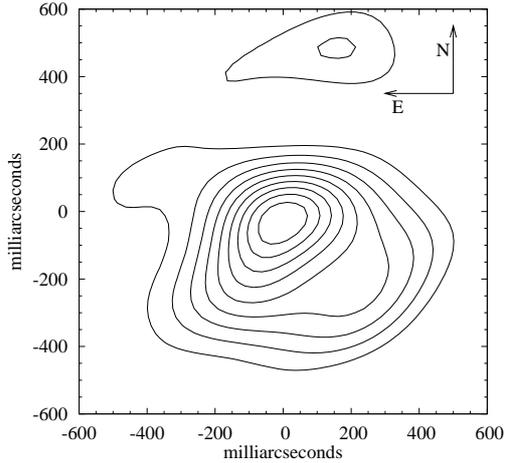}
  \caption{T-ReCS image of IRS\,9A at 11.7\,$\upmu$m. North is up, and east is
    to the left. Reconstruction from the aperture masking data using the
    Maximum Entropy Method (MEM). Contour levels are the same as in 
    Fig.~\ref{f_acq}.}
  \label{f_Gemini_image}
\end{figure}

In order to derive an image from the set of amplitudes and phases, we employed
the VLBMEM imaging software \citep{sivia1987}, which uses a Maximum Entropy
algorithm. The resulting image is shown in Fig.~\ref{f_Gemini_image}. It agrees
well with the overall appearance of our MIDI acquisition images, again showing
the emission of IRS\,9A to be quite extended and slightly asymmetric. 

\subsection{Spitzer} \label{obs_Spitzer}

\begin{figure}
  \includegraphics[width=1.0\columnwidth]{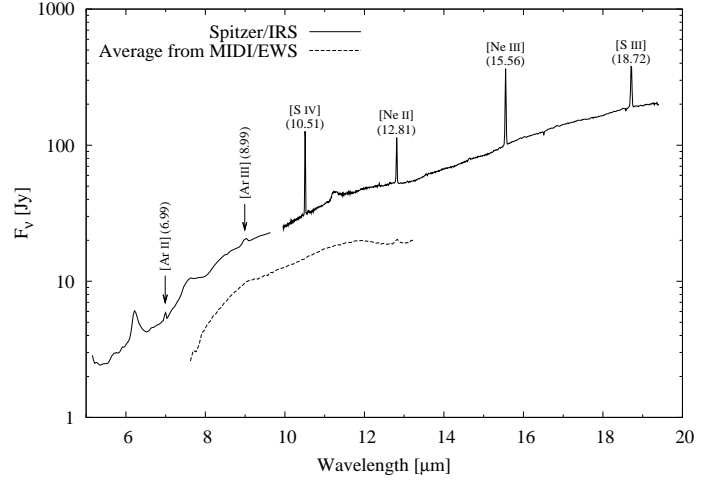}
  \caption{The spectrum of IRS\,9A from Spitzer (solid lines) 
           and MIDI (dotted line). The numerous
           forbidden emission lines are labelled together with their central
           wavelengths. There are also several PAH features visible, with the 
	   most prominent ones lying at 6.2, 7.6 and 11.2\,$\upmu$m.
	   The small gap is due to saturation of Spitzer's SL
           module for wavelengths larger than 9.7$\,\upmu$m. The data from the
           LH module (18.7\,--\,37.2$\,\upmu$m) suffered from the same effect
           and are not shown \citep[see also][]{lebouteiller2008}.}
  \label{f_SED_Spitzer}
\end{figure}

In Fig.~\ref{f_SED_Spitzer}, we show the SED of IRS\,9A measured
by \citet{lebouteiller2007, lebouteiller2008} with the Spitzer Space
Telescope in comparison with the MIDI spectrum.  Since Spitzer's spatial
resolution is about a factor of 10 worse compared to the (MACAO-assisted)
VLT, the IRS spectrograph aboard Spitzer collects considerably more flux,
about twice that of MIDI in the middle of the N band.  There is no
hint of the [S\,{\footnotesize IV}] line in our MIDI data, although
its line flux is more than two times higher than the one from the
[Ne\,{\footnotesize II}] line (see \citet{lebouteiller2008}) which we
did barely detect with MIDI (see Fig.~\ref{f_EWS_spectrum}). It is
possible that the region emitting the [S\,{\footnotesize IV}] line is
more extended than the MIDI field of view.
We will use the Spitzer SED as an additional constraint for our models in 
Sect.~\ref{analysisandmodels}.

\section{Model fitting and analysis} \label{analysisandmodels}

In this section, we first try to model the circumstellar structure
of NGC\,3603 IRS\,9A in the context of current star formation scenarios
by using such elements as rings, disks, and envelopes. We then describe
our efforts of testing physical models which require radiative transfer
computations to produce images of the structure for comparison with
the available visibility and photometric data. However, given that the
environment of young and very luminous stars is complex, our models
can only represent a first step on the way of developing an adequate
physical model of IRS 9A. Also the paramerters of these models may not
be well constrained given their large number.

\subsection{Simple geometrical models} \label{geometricalmodels}

\begin{figure}
  \includegraphics[width=1.0\columnwidth]{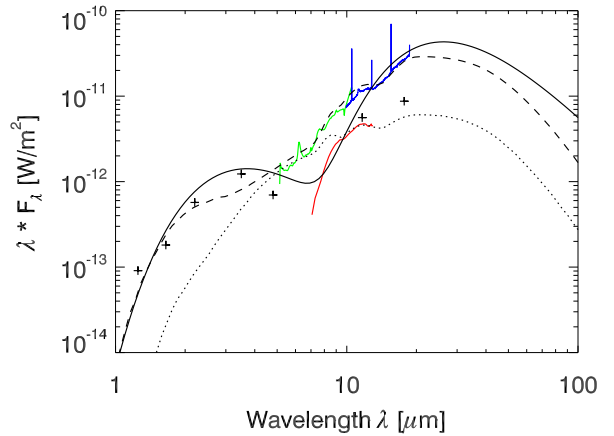}
  \caption{The SED of the geometrical model (solid line) with the observations
    from Spitzer (green and blue solid lines), MIDI (red solid line), and
    NIR/MIR photometry (plus symbols) published by \citet{2003A&A...404..255N}.
    The two components of the model have the same flux at 8 microns.
    The dashed and dotted lines correspond to the physical model described in 
    Section~\ref{Diskmodels}, with aperture radii of $3\arcsec$ (Spitzer) and 
    $0.3\arcsec$ (MIDI), respectively.}
  \label{f_ringgaussian}
\end{figure}

It is useful to identify structural components of the source and
derive the parameters of such simple geometrical models like sizes and
orientations. Beginning with the visibilities measured with T-ReCS in a
single filter (hence no wavelength dependence of the source structure need
to be considered), we first fit circular Gaussian and uniform disk models
with diameters of 311\,mas (FWHM) and 495\,mas, respectively. The reduced
$\chi^2$ (henceforth denoted with $\chi^2_{\rm r}$) of the Gaussian model
can be further reduced to 2.3 by allowing an ellipsoid, with a major
axis of 329\,mas, PA of 100$\degr$ east of north, and a minor axis of
282\,mas (axial ratio 0.86, compare with values from Table~\ref{t_acq}).
This component would be completely resolved by MIDI even on the shortest
baselines available, and therefore a second component of small angular
scale needs to be added to the geometrical model in order to explain
the MIDI detection. After trying unsuccessfully to fit the data with an
additional uniform disk, we adopt a ring-shaped component instead because
its Fourier transform exhibits a larger secondary maximum (beyond the
null) than a uniform disk. The ring, which could be associated with the
inner hot regions of a disk with a central hole, is initially assigned a
blackbody temperature of 1500\,K (equal to the typical dust sublimation
temperature) and a fractional width of 1/4 of its radius.
We assume here, before a more realistic physical model
is applied later on, that the two components are co-located and do not
obscure each other. In order to have the flux of the Gaussian component
decrease towards the blue end of the  N-band so that the correlated flux
of the ring begins to dominate here the visibility measured by MIDI
(see Fig.~\ref{f_ringgaussian}), we assign initially a low blackbody
temperature of 250\,K to the extended Gaussian component based on
a spectral decomposition by \citet{2003A&A...404..255N}. This model
(``geometrical model'') indeed results in a good fit ($\chi^2_{\rm
r}=1.6$) of all visibility data (see Figs.~\ref{f_MIA_V1}
and \ref{f_MIA_V2}, as well as Figs.~\ref{f_Gemini_uvr} and
\ref{f_Gemini_blpa_1}), with the best-fit parameter values given in 
Table~\ref{t_geometrical}.
The inclination of the polar axis of the
ring against the line of sight is 71 degrees. 
The fit to the SED is shown in Fig.~\ref{f_ringgaussian}.

\setlength{\tabcolsep}{1mm}
\begin{table}
  \caption{The parameters of the geometrical model.
	   }
  \centering
  \begin{tabular}{ll|rc}
    \hline\hline
    Parameter		 	& Unit		&  Value  \\ \hline
	Ring major axis		& [mas]		& 57 \\
	Major axis PA		& [$\degr$]	& 105$^a$ \\
	Ratio minor/major axis  &		& 0.32 \\
	Fractional ring width		& 		& 0.31 \\
	Ring effective temperature	& [K]		& 1000 \\
	Envelope major axis FWHM& [mas]		& 350 \\
	Major axis PA & [$\degr$]	& 105$^a$ \\
	Ratio minor/major axis &		& 0.86 \\
	Envelope eff. temperature & [K]		& 140 \\ \hline
$^a$: Forced to be the same value.
  \end{tabular}
  \label{t_geometrical}
\end{table}

There is no clear evidence for a companion of IRS\,9A, and we can
use the geometrical model to estimate at what level we can exclude
a companion.  We adopt the companion to be a ring like the primary so
that the binary model asymptotically approaches the geometrical model
for small separations when the individual component fluxes are adjusted
(given a magnitude difference) to yield a constant total flux.  For the
limiting $\chi^2_{\rm r}$, we adopted 1.7 which results in recognizable
deviations from the data.  Almost all models below this limit have a
magnitude difference of 3 or larger, and therefore we can exclude the
possibility that a companion of more than about 0.5\,Jy exists between
30\,mas (the resolution limit) and 500\,mas (our search area) away from
the primary. Due to the fact that uncorrelated flux dominates at the
longer wavelengths where none of the models could therefore produce any
significant signatures, we cannot place more stringent limits.

\subsection{Spherical dust shell models} \label{DUSTYmodels}

We now interpret the geometrical model we have studied as spherical dust
shells with the innermost one corresponding to the ring component
and with the star at their center.  For the calculation of the radiative
transfer we use the DUSTY code \citep{ivezic1997,ivezic1999a,ivezic1999b}
which adopts spherical symmetry. It produces detailed spectra and images
at specified wavelengths, and the latter can be used to calculate the
visibilities of the models that correspond to the observational setup of
T-ReCS or MIDI. These types of models have been used quite successfully
by \citet{dewit2007} and \citet{dewit2009} for their MIDI observations
of W33A and resolved $24.5\,\upmu$m emission of MYSOs, respectively.

\begin{figure}
  \includegraphics[width=1.0\columnwidth]{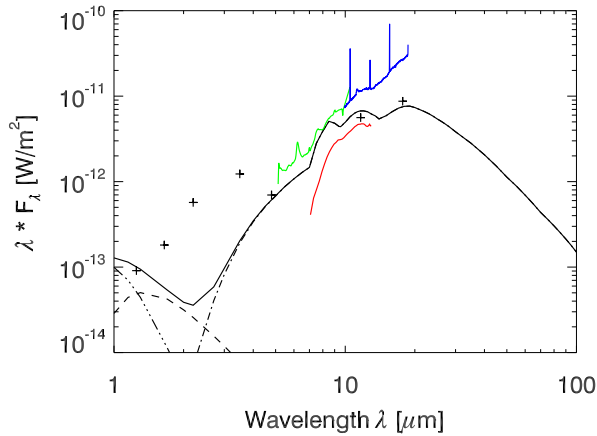}
  \caption{Comparison of the SED of our DUSTY model with the observations from
    Spitzer, MIDI, and the NIR/MIR photometry (plus symbols) from
    Fig.~\ref{f_ringgaussian}. The dashed line corresponds to the attenuated
    input radiation, the dash-dotted line to the dust radiation, the
    dash-double-dotted line to the scattered emission.}
  \label{f_DUSTY_SEDs}
\end{figure}

For the central star we adopted an effective temperature of 22\,000\,K,
and for the shells a density, $\rho$, proportional to $r^{-2}$, where
$r$ is radius of the shell. For the dust composition we used standard
astronomical silicates.  The most important free parameters to adjust
were the temperature, $T_{\rm s}$, of the inner edge of the shells,
and its opacity, $\tau$.  In general, a lower $T_{\rm s}$ increased
the size of the shells and thus lowered the visibility amplitudes,
while a larger opacity had a similar effect yet introduced a strong
silicate absorption feature in the spectrum. Using the T-ReCS data
and the SED as constraints, we settled on a model with $T_{\rm s} =
500$\,K and $\tau_{10\upmu\rm m} = 1.5$, corresponding to $A_V=29$
mag \citep{1985ApJ...288..618R}. The fit to the T-ReCS visibilities
was indistinguishable from the (circular) Gaussian model, while the
fit to the SED is shown in Fig.~\ref{f_DUSTY_SEDs}. DUSTY returns the
radius of the innermost shell, $r_{\rm in}$ (where $T=T_{\rm s}$), which
scales proportional with the square root of the luminosity, $L^{1/2}$.
Adopting a luminosity of IRS\,9A of 230\,000\,$L_{\sun}$, we derive
$r_{\rm s}=170$\,mas. The (dust) mass of the shell can be derived from
the opacity $\tau$, the absorption coefficient $\kappa$, and the ratio
of the outer edge of the shell to $r_{\rm in}$, $Y$, as follows,

\begin{eqnarray}
\tau & = \int_{r_{\rm in}}^{r_{\rm out}} \kappa \rho \,ds 
     & = \rho_0 \kappa \frac{r_{\rm in}}{r_{\rm out}} (r_{\rm out}-r_{\rm in}),
\end{eqnarray}

\begin{eqnarray}
M & = \int_{r_{\rm in}}^{r_{\rm out}} \rho 4\pi r^2 \,dr 
  & = 4\pi\frac{\tau}{\kappa}r^2_{\rm in} Y,
\end{eqnarray}
giving a dust mass of about 1$M_{\sun}$.

We thus find that it is possible to devise a physical model for the T-ReCS
data set, but when we apply that model to the observations from MIDI, the
agreement is very poor because the emission is completely resolved spatially.
This also applies vice-versa, that is we are able to find models with
DUSTY which can reproduce the visibilities measured by MIDI (albeit showing
larger deviations), but these models are in turn a very poor fit to the data
from T-ReCS. In order to find suitable models that are able to account for all
our observational data, we conducted a grid search varying all parameters
except for $T_{\rm s}$, $Y$ and the temperature of the central star. However,
none of the resulting models could accomplish this task.

It thus seems obvious that spherical dust distributions are not compatible with
the circumstellar structure of IRS\,9A. Therefore, in the next section, we
examine the effects of non-spherical dust distributions on the visibilities
and the SED.

\subsection{Disk models} \label{Diskmodels}

Within the framework of models for YSOs, the most obvious approach for
a non-spherical density distribution is given by the flattened structure
of a circumstellar disk embedded in an envelope. If seen
under low (near face-on) or intermediate inclinations of the disk axis
to the line of sight, a disk opens a more or less direct optical path to
the central object and the hot inner disk regions (the hot ring in our
geometrical model), while at the same time maintaining an extended dust
distribution (the cool Gaussian component of the geometrical model).
Models involving disks have already been used widely for interpreting
SEDs and interferometric observations of T~Tauri and Herbig Ae/Be
stars. For example, the visibilities predicted by models computed for
Herbig Ae/Be stars by \citet{2004A&A...423..537L} look similar to our
MIDI visibilities. In this section we report on our first attempt
to fit such models to our data.

We adopt the disk-envelope model of \citet{2003ApJ...591.1049W}
which has been used by \citet{robitaille2006} to compute
a large grid of model SED, made available through a web
server\footnote{\url{http://caravan.astro.wisc.edu/protostars/}}.
Following \citet{2009A&A...505..655L}, we select only those models fitting
the Spitzer-SED of IRS9A and compute their images. To compute the model
visibilities, we apodized the images at scales larger than the PSF of an
8\,m telescope in the case of MIDI, while due to the comparitively small
subapertures in T-ReCS, this apodization was not necessary. Thus we find
that the model 3012790 inclined at $85^\circ$ to the line of sight (nearly
edge on) and with the disk roughly oriented East-West provides the best
fit of the ten selected models. The parameters of this model are listed
in Table \ref{t_YSOgrid}.  The disk-envelope model 3012790 includes a
significant accretion rate, for which however no independent constraint
exists so far.  The fits of this model to the visibilities are shown
in Figs.~\ref{f_MIA_V1}, \ref{f_MIA_V2}, and Fig.~\ref{f_Gemini_uvr},
while the SED is shown in Fig.~\ref{f_ringgaussian}.

\setlength{\tabcolsep}{1mm}
\begin{table}
  \caption{The parameters of the Robitaille
           disk-envelope model 3012790.
	   }
  \centering
  \begin{tabular}{ll|rc}
    \hline\hline
    Parameter		 	& Unit		&  Value  \\ \hline
    Stellar mass                & [$M_{\sun}$]  &  25     \\
    Stellar radius              & [$R_{\sun}$]  &  6.5    \\
    Effective temperature       & [K]           &  38000  \\
    Luminosity                  & [L$_{\sun}$]  &  92000  \\
    Inner disk/envelope radius  & [AU]          &  25     \\
    Outer disk radius           & [AU]          &  94     \\
    Outer envelope radius       & [AU]          &  100000 \\
    Disk dust mass              & [M$_{\sun}$]  &  0.005  \\
    Envelope dust mass          & [M$_{\sun}$]  &  0.9    \\
    Inclination                 & [$\degr$]     &  85     \\
    Disk flaring power, $\beta$ &               &  1.2    \\
    Disk scale height           & [AU]          &  9      \\
    Cavity cone angle           & [$\degr$]     &  29     \\ \hline
  \end{tabular}
  \label{t_YSOgrid}
\end{table}

\section{Discussion} \label{discussion}

Considering the MIDI observations, the most important feature to reproduce
by any model is the steep rise of the visibility towards 8 microns,
indicating the emergence of warm inner dusty regions. The disk-envelope
model is able to reproduce this feature about as well as the geometrical
model, but the envelope is too large considering the low predicted
visibilities for the T-ReCS observations.

The T-ReCS image of IRS\,9A (Fig.~\ref{f_Gemini_image})
resembles the synthetic images of class I sources computed by
\citet{2003ApJ...598.1079W} (their Fig.~11a) and
indicates the presence of a bipolar cavity facing the observer and
pointing towards the South to South-West. Ionizing
stellar radiation escaping through the cavity has created an HII region as
indicated by the detection of the [Ne\,{\footnotesize II}] line by Spitzer
(see Fig.~\ref{f_SED_Spitzer}). This morphology of the circumstellar
dust (and gas) distribution is consistent with the interpretation of
NACO images of NGC\,3603 IRS\,9A by \citet{2008JPhCS.131a2025N} which
show an envelope with a major-to-minor axis ratio of about 2:1 at a
position angle of $82^\circ$. (The NACO observations also confirm the
reality of the faint feature in our T-ReCS image about 0.6 arcseconds to
the north of IRS\,9A.)  Due to the limited dynamic range of our images,
we cannot confirm the ablation of dust and gas towards the south-east,
i.e. away from the NGC\,3603 cluster center, as seen in the NACO images.

A lower mass estimate of $0.1\,M_{\sun}$ for the hot dust and gas
bound to IRS9A was derived by \citet{2003A&A...404..255N}
based on the assumption that the MIR emission is optically thin.
For the envelope mass, an estimate of $1\,M_{\sun}$ is derived by
the same author based on an estimate of the intrinsic extinction of
$A_V=10-15$, much lower than our estimate based on DUSTY. However,
\citet{2003A&A...404..255N} remarked that a radio flux measurement of
NGC\,3603 IRS\,9A yielded unexpectedly high values, which would argue
in favour of a higher envelope mass estimate. A final answer cannot be
given until a sub-mm flux measurement can be obtained.

\section{Conclusions} \label{conclusions}

We gave a detailed account of our efforts to model the observed spatial
and spectral data of the MYSO candidate NGC\,3603 IRS\,9A. As a first
result, we find that the observed visibilities are not compatible with
a binary nature of IRS\,9A. However, given the limits in sensitivity
($\approx 0.5\,$Jy; the correlated flux on longer baselines would have
been too small to detect) and in angular resolution (down to 30\,mas) of
our interferometric data, we cannot exclude the presence of a companion
with either significantly lower mid-infrared flux (less than 10\,\%)
and\,/\,or an extremely tight binary with a separation of less than
about 100\,AU.

We have shown that a spherical dust distribution is inconsistent with
the data and that therefore a flattened distribution is more likely.
A geometrical model consisting of a warm ring and a cool extended
component can reproduce all observations fairly well, but requires
the ad-hoc assumption of a direct line of sight into the inner regions
of the warm dust.  The ring has a temperature 500 K less than the dust
sublimation temperature (1500 K), probably a consequence of the fact that
it represents not only the very inner edge of a disk, but rather the inner
regions of it.  Physical disk-envelope models provide a self-consistent
connection between the hot inner and the cooler outer regions and allow
the former to be seen more directly. However, no single model was able
to fit all observations simultaneously. Since it could be expected from
the early evolutionary stage of class I sources compared to the well
developed disks of the class II sources (adopting the classification
scheme more strictly developed for lower mass objects), that the dust
structures are more complex, e.g. clumpy or showing disk distortions,
our data are not sufficient to prove the existence of a disk.

Comparing NGC\,3603 IRS\,9A to other MYSOs for which MIDI
interferometry was published, e.g.\@ W33A \citep{dewit2007}, M8E-IR
\citep{2009A&A...505..655L}, and MWC\,297 \citep{acke2008}, we
find that all of them exhibit rather different properties. These depend
mostly on the amount of circumstellar material present, as reflected
in the estimated visual extinctions which range from $A_{\rm V}=22$
to $A_{\rm V}=100$. The source with the highest extinction, W33A,
displays a deep silicate absorption feature. The measured visibilities
were similar to the ones we measured between 8 and 9\,$\upmu$m on the
same baseline, translating into a Gaussian FWHM of 120\,AU for W33A
(distance of 3.8\,kpc). A significant difference here to NGC\,3603 IRS\,9A
is that the latter is completely resolved by MIDI at longer wavelengths
due to the envelope detected with T-ReCS. Thus, W33A seems a lot more
compact, and was therefore successfully modelled with a single DUSTY
envelope. The model required a rather low effective stellar temperature
for a $10\,M_{\sun}$ star (only 10\,000\,K), possibly indicative of a
star swollen due to accretion. The same effect was observed in M8E-IR,
which was modelled with an envelope and disk surrounding a rather
cool (5\,000\,K) star with a mass of 13.5\,$M_{\sun}$. M8E-IR has a much
weaker silicate absorption feature, and appears to be rather compact
(45 - 100\,AU). The existence of a disk in the envelope could not be
ascertainted. Yet again, the difference to NGC\,3603 IRS\,9A lies in
the fact that the visibility of M8E-IR is only weakly dependent on
wavelength, indicating that a flattened structure is not necessary to
explain them. Finally, the observations of MWC\,297 could not be modeled
with a physical disk model at all. This star is a young 10\,$M_{\sun}$
Be star with a rather low extinction. Only indirect arguments favoured
the disk geometry for the circumstellar material.

\begin{acknowledgements}
We are grateful to Vianney Lebouteiller for kindly providing the Spitzer
spectrum of IRS\,9A. 
We also thank Mario van den Ancker for sharing his insight
during several discussions of this work. S.V.\ acknowledges support from DFG
via SFB\,439 and from the University of Kiel, Germany. J.D.M.\ acknowledges
support from NASA NNG05GI80G, NSF-AST0352728. Some observations contained
herein were obtained at the Gemini Observatory, which is operated by the
Association of Universities for Research in Astronomy, Inc., under a
cooperative agreement with the NSF on behalf of the Gemini partnership:
the National Science Foundation (United States), the Science and Technology
Facilities Council (United Kingdom), the National Research Council (Canada),
CONICYT (Chile), the Australian Research Council (Australia), Minist\'erio da
Ci\^encia e Tecnologia (Brazil) and Ministerio de Ciencia, Tecnolog\'ia e
Innovaci\'on Productiva (Argentina). This research has made use of the SIMBAD
database, operated at CDS, Strasbourg, France.
\end{acknowledgements}

\bibliographystyle{aa}
\bibliography{13546}

\end{document}